```
program Alpha;  {  Calculation of the fine structure constant  } {I FOSAE}
                {       using the result of my discovery        } {N AFCNN}
{$N+} uses crt; { - Theory of the electron charge essence -     } {G C IDG}
var                                                                {. UNE I}
    i,j,k,kmax : integer;                                          {  LUNPN}
    b : array [1..30] of extended;                                 {J TCCHE}
    k1,p,x,a0,a1,a2,ac : extended;                                 {. YLEYE}
begin                                                              {P  E SR}
    window(1,1,80,25);                                             {A  A II}
    textattr := 0; clrscr;                                         {V  R CN}
    textattr := 7;                                                 {E    AG}
    window(6,2,71,25); write(                                      {L    L }
```

The fine structure constant: $\alpha = 1/137.0360037000$

$$\alpha o = \int_0^{\pi/2} \int_1^2 \left[1+\left(1-\frac{\mu}{\sin^2\mu}\right)\cdot\frac{2}{r^2}\right]\cdot\frac{\sin\mu}{2\cdot r^2}\, dr\, d\mu$$

$$\alpha c = \alpha\cdot[1+\alpha/3/(1-2\cdot\alpha)]/\sqrt{1-\alpha^2} \qquad n = 3/2$$

$$\alpha = \alpha o\cdot\sqrt{2/n-\alpha c}\cdot\sqrt{n/2-\alpha c}$$

$$R = \int_0^{\pi/2} x/\sin(x)\ dx$$

$$\alpha o = (13-7\cdot R)/24$$

$$\alpha = e^2\cdot\mu o\cdot c/2h$$

Jaroslav Pavel
Brdickova 1914
155 00 Praha 5
Czech Republic

(c) Praha 1999

E-m: pavel@troja.fjfi.cvut.cz

e is the electron charge , h is the Planck constant
$\mu o = 4\pi\cdot 1E{-}07$ H/m is the permeability of vacuum
c = 299792458 m/s is the speed of light in vacuum
');

```
{a}  kmax := 30; b[1] := 1/6;                                      {DOE C1V}
     for i := 2 to kmax do begin p := 2*i; x := 0; k1 := 1;        {EFL Z8 }
     for j := 1 to i-1 do begin k := 2*j; k1 := k1*k*(k-1);        {P E EOH}
     x := x+p*b[j]/k1; p := p*(2*i-k+1)*(2*i-k); end;              {APC COO}
     b[i] := 1/2-1/(2*i+1)-x;   { the numbers of Bernoulli } end;  {RHT H0L}
{b}  x := (13-7*pi/2)/24; k1 := 1; for k := 1 to kmax do           {TYR  E }
     begin p := exp((2*k-1)*ln(2)); k1 := k1*2*k*(2*k+1);          {MSO RPS}
     x := x-7*pi*abs(b[k])*(p-1)*exp(2*k*ln(pi))/(48*p*k1); end;   {EIN ERO}
{c}  a0 := x; a1 := a0; repeat a2 := a1;                           {NCI PAV}
     ac := a2*(1+a2/3/(1-2*a2)/sqrt(1-sqr(a2)));                   {TAC UHI}
     a1 := a0*sqrt((3/4-ac)*(4/3-ac)); until abs(a1-a2) = 0;       { LS BAC}
{d}  gotoxy(33,2); textcolor(white); write('α = 1/',1/a1:13:10);   {   L K}
     gotoxy(35,17); textcolor(yellow);                             {   I8A}
     write('Calculation of the constant α');                      {   C C}
     window(80,25,80,25); repeat until KeyPressed;                 {     H}
     window(1,1,80,25); textattr := 7; clrscr;                     {     }
end. { This program has been written in Turbo Pascal language. }  {     2}
```